\documentclass[12pt]{article}
\usepackage{graphicx}

\def\bm#1{{\mbox{\boldmath $#1$}}}
\def\bms#1{{\mbox{\rm\scriptsize\it\bf#1}}}

\def\e{\epsilon}

\def\t{\tau}
\def\D{\Delta}
\def\del{\partial}

\def\s{\sigma}

\def\bm#1{{\mbox{\boldmath $#1$}}}
\def\bms#1{{\mbox{\rm \scriptsize \bf #1}}}

\def\dd{{\mbox{d}}}

\begin{document}

\title{\bf Thermoelectric Power of Insulators and Reconsideration of Kelvin's Relations at Low Temperatures}

\author{T Saso\\
Department of Physics, Faculty of Science, Saitama University,\\
Shimo-Ohkubo 255, Saitama-City, Saitama 338-8570, Japan
}
\date{}
\maketitle

\begin{abstract}
Thermoelectric effects in Kondo insulators are attracting interests because of the emerging possibility of developping better thermoelectric materials for a portable refrigerator without liquid coolant.  In this article, the theory of thermoelectric effects are reinvestigated for insulators or semiconductors at low temperatures.  It is found that the famous relations established by Lord Kelvin for metals in 1851 must be modified for insulators in order to be consistent with the third law of the thermodynamics.  Effects of strong correlation are discussed.
\end{abstract}

\section{Introduction}
Recently, some of the strongly correlated materials are attracting renewed interests because of a possibility for a new and efficient themoelectric device.\cite{Mahan98}  Especially, the compounds called as the Kondo insulators\cite{Aeppli92} show rather large thermopower at low temperatures, being suitable for a refrigerator without coolant in the lower temperature range.  It seems, however, that the theory for the thermoelectric effects has not yet been fully developped for the insulators and for the systems with strong correlation.

In this article, we will focus on the theory of the thermoelectric effect in insulators and point out for the first time that the famous Kelvin's relations, which were analyzed and established by Lord Kelvin\cite{Thomson1851} 150 years ago, must be modified in the case of the insulators at low temperature limit.  Next, we will discuss effects of many-body interactions on the thermoelectric coefficients and give a formula for the heat current.  We will also explain how the behaviour of the thermopower due to the non-interacting electrons or holes in ordinary semiconductors might be changed in the Kondo insulators with strong correlations at low temperatures.

\section{Reconsideration of Kelvin's relation in insulators}
Lord Kelvin\cite{Thomson1851,Wilson65} analyzed a system depicted in Fig. 1 which consists of the metals a and b.  The temperature difference between B and C is set equal to $\D T$.  (Note that the direction of the gradient is opposite to that in ref.\cite{Wilson65})  A unit charge is quasi-statically moved along the path ABCDA.  Thereby the Peltier heat $\pi_{ab}(T+\D T)$ is emitted at B and $\pi_{ab}(T)$ is absorbed at C.  The Thompson heat $\tau_b\Delta T$ is emitted between BC and $\tau_a\D T$ is absorbed between AB and CD.  On transfering the charge from D to A the work $S_{ab}\D T$ is done outwards due to the thermoelectric voltage.  Thus we can set up the following equations to express the first and the second laws of the thermodynamics:
\begin{equation}
  -S_{ab}(T)\D T + \pi_{ab}(T+\D T)-\pi_{ab}(T) + (\t_b-\t_a)\D T = 0, \label{eq:Kelvin1}
\end{equation}
\begin{equation}
  \frac{\pi_{ab}(T+\D T)}{T+\D T}-\frac{\pi_{ab}(T)}{T} + \frac{\t_b-\t_a}{T}\D T = 0,
\end{equation}
and derive the famous Kelvin's relations:
\begin{equation}
  S_{ab}(T) = \frac{\pi_{ab}(T)}{T}, \quad
  S_{ab}(T) = \int_0^T \frac{\t_a(T')-\t_b(T')}{T'}\dd T'
\label{eq:Onsager}
\end{equation}
where the Seebeck coefficient is related to $\Theta_{ab}$ by $S_{ab}=\Theta_{ab}/\D T$.
It is well known that the absolute Seebeck coefficient is given by $S \simeq -(E_c-\mu)/|e|T$ for the semiconductors at low temperatures when the carriers are electron-like, hence $S_{ab}=S_a-S_b \simeq (E_c^b-E_c^a)/|e|T$. Here, $E_c$ and $\mu$ denote the conduction band edge and the chemical potential, respectively.  However, according to the first equation in (\ref{eq:Onsager}), $\pi_{ab}(T\rightarrow 0) \rightarrow (E_c^b-E_c^a)/|e|$, whereas $S(T)$ must vanish at $T\rightarrow 0$ due to the second equation in (\ref{eq:Onsager}).  The former means that one can remove finite amount of heat from the body at absolute zero temperature, violating the third law of the thermodynamics.

%%%%%%%%%%%%%%%%%%%%%%%%%% Figure %%%%%%%%%%%%%%%%%%%%%%%%%%%%%%%%%%%%%
\begin{figure}[t]
\begin{center}
\includegraphics[width=12.0cm]{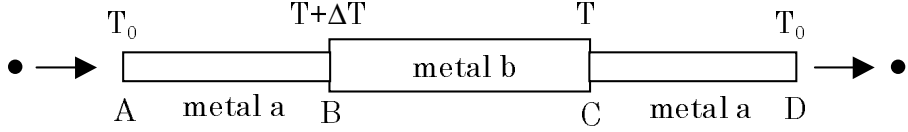}
\end{center}
\caption{A schematic figure to derive Kelvin's relations.  A unit charge is quasi-statically moved along the path ABCDA to prove Kelvin's relation.}
\end{figure}
%%%%%%%%%%%%%%%%%%%%%%%%%%%%%%%%%%%%%%%%%%%%%%%%%%%%%%%%%%%%%%%%%%%%%%%

This contradiction seems to have occured because we have neglected the work necessary to move a negative charge from the bottom of the conduction band at $E_c^a$ to the region in the middle with the band edge $E_c^b$ when $E_c^b > E_c^a$ and $T \rightarrow 0$, as shown in Fig. 2.  In this case, we have to add a term $W=(E_c^b-E_c^a)/|e|$ in eq. (\ref{eq:Kelvin1}), which leads to the modified Kelvin's relations,
\begin{equation}
  S_{ab}(T) = \frac{\pi_{ab}(T)}{T} + \frac{E_c^b-E_c^a}{|e|T}, \label{eq:neweq1}
\end{equation}
\begin{equation}
  S_{ab}(T) = \int_0^T \frac{\t_a(T')-\t_b(T')}{T'}\dd T' + \frac{E_c^b-E_c^a}{|e|T}. \label{eq:neweq2}
\end{equation}
In deriving these equations, we have replaced $T \rightarrow 0$ and $T+\D T \rightarrow T$.  Thus, the diverging behavior of $S$ at low temperature limit can be understood as due to the necessary `work' $W$ to move a charge from the lower to the higher bottom of the conduction bands at low temperature limit.  The heat which can be removed ($Q=\pi_{ab}(T)+(\tau_a-\tau_b)T$) must vanish at $T \rightarrow 0$ in this representation, and the finite value $\pi_{ab} \rightarrow (E_c^b-E_c^a)/|e|$ is an artifact of regarding the work to be done as the heat to be removed.  (See Fig.\ref{fig:cooling} to understand how the cooling apparatus works as a thermodynamic cycle.)  Onsager's relation (eq.(\ref{eq:Onsager})) does not hold, since the linear response theory is not applicable in the present case at $T \rightarrow 0$.  Thus, $S_{ab}(T) \rightarrow (E_c^b-E_c^a)/|e|T$ still holds but $\pi_{ab}(T) \rightarrow 0$ at $T \rightarrow 0$.  Note that the equations (\ref{eq:neweq1}) and (\ref{eq:neweq2}) hold only at the low temperature limit and a more general expressions at arbitrary temperature are still to be found.  Thereby, the inclusion of the nonequilibrium effects will be necessary.

%%%%%%%%%%%%%%%%%%%%%%%%%% Figure %%%%%%%%%%%%%%%%%%%%%%%%%%%%%%%%%%%%%
\begin{figure}[t]
\begin{center}
\includegraphics[width=8.0cm]{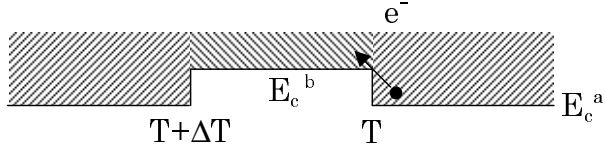}
\end{center}
\caption{The hatched regions indicate the conduction bands, whereas the region beneath them are the gaps.  Motion of a negative charge through the junction between the two semiconductors with the conduction band bottom $E_c^a$ and $E_c^b$, when a positive charge is moved from left to right.}
\end{figure}
%%%%%%%%%%%%%%%%%%%%%%%%%%%%%%%%%%%%%%%%%%%%%%%%%%%%%%%%%%%%%%%%%%%%%%%

%%%%%%%%%%%%%%%%%%%%%%%%%% Figure %%%%%%%%%%%%%%%%%%%%%%%%%%%%%%%%%%%%%
\begin{figure}[t]
\begin{center}
\includegraphics[width=3.5cm]{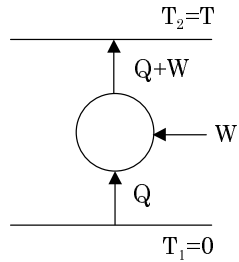}
\end{center}
\caption{A schematic figure of a thermodynamic apparatus for cooling.}
\label{fig:cooling}
\end{figure}
%%%%%%%%%%%%%%%%%%%%%%%%%%%%%%%%%%%%%%%%%%%%%%%%%%%%%%%%%%%%%%%%%%%%%%%

\section{Seebeck coefficient and many-body effect}
A general theory for the thermoelectric transport has been developped by Mahan,\cite{Jonson90} and recently by Kontani\cite{Kontani01} further, based on the Fermi liquid theory.  It should be noted that the heat conveyed by the carriers is not equal to $\e_\bms{k}-\mu$ but a correction due to the interaction exists.  For example, the heat current is given by
\begin{equation}
  \bm{j}_Q = \sum _{\bms{k}\s}(\e_\bms{k}-\mu)\bm{v}_\bms{k} c^+_{\bms{k}\s}c_{\bms{k}\s} 
   + U \sum_{\bms{k}\s} \bm{v}_\bms{k}\left( c^+_{i\s}c_{j\s} \frac{n_{i-\s}+n_{j-\s}}{2}\right)_\bms{k},
  \label{eq:heat-current}
\end{equation}
for the Hubbard model, where $(\cdots)_\bms{k}$ denotes the Fourier transformation from $i,j$ to $\bm{k}$.  The Seebeck coefficient, however, can be expressed by the following simple formula
\begin{equation}
  S(T) = -\frac{1}{|e|T} \frac{\int\dd\e (\e-\mu) L(\e) \left(-\frac{\del f}{\del \e}\right)}{\int\dd\e L(\e) \left(-\frac{\del f}{\del \e}\right)}
  \label{eq:Seebeck-Boltzmann}
\end{equation}
with $L(\e)=\sum_\bms{k}v_{x\bms{k}}^2\left[ \mbox{Im} G(\bm{k},\e)\right]^2/\pi N$ when the vertex corrections (VC's) can be neglected, as would be the case for the heavy-fermion compounds.  VC's cannot be neglected for high $T_c$ cuprates.\cite{Kontani01}  It should be noted that the second term in eq. (\ref{eq:heat-current}) due to the interaction does not appear in the factor $(\e-\mu)$ in eq. (\ref{eq:Seebeck-Boltzmann}).  It is because all the interaction effects were absorbed in the Green's function (and in the VC's).

The electronic states of the Kondo insulators can be described most simply by the periodic Anderson model at half-filling.  (A more realistic model will be discussed in \cite{Saso02b}.)  The Coulomb interaction between f-electrons produces a finite imaginary part of the self-energy at finite temperatures.  This makes the quasi-particle density of states finite within the gap at $T>0$.  Therefore, the Seebeck coefficient of the Kondo insulators becomes metal-like $S(T) \propto T$ at low but finite temperatures,\cite{Saso02} whereas $S(T) \propto T^{-1}$ at higher temperatures.  It could be the case that $S$ vanishes as an activation type $S(T) \propto \exp(-E_g/k_BT)$ at the low temperature limit.  These predictions are qualitatively consistent with the observed behaviors,\cite{Iga01,Sales94} but the effect of the nonstoichiometry may not be neglected.  Both interpretations seem possible for analyzing the experimental results.

\section{Summary}
  The thermoelectric effect in insulators or semiconductors was reinvestigated theoretically.  It was found that the famous relations established by Lord Kelvin for the thermoelectric coefficients must be modified at low temperature limit in order to be consistent with the third law of the thermodynamics.  Many-body effect was also discussed and it was noted that $S(T) \propto T$ or $\propto \exp(-E_g/k_BT)$ at low temperatures due to the interaction effect.  To elucidate the low temperature behavior, it may be necessary to consider the nonequilibrium effect of coupled electron and phonon systems.

%%%%%%%%%%%%%%%%%%%%% Acknowledgement %%%%%%%%%%%%%%%%%%%%%%%%%%%%%
%\section*{Acknowledgements}
This work is supported by the Grant-in-Aid for Scientific Research
from the Ministry of Education, Science, Sports, Culture, and
Technology.

%%%%%%%%%%%%%%%%%%%%% References %%%%%%%%%%%%%%%%%%%%%%%%%%%%%%%

\end{document}